\documentclass{revtex4}
\usepackage{graphicx}
\usepackage{latexsym,amssymb}
\begin{document}

\title{Peeling from a  patterned thin elastic film}
\author{Animangsu Ghatak $^{\sharp}$ \footnotemark[1], L.
Mahadevan $^{\star}$\footnotemark[1]\footnotemark[2], Jun Young
Chung $^{\sharp}$, Manoj K. Chaudhury $^{\sharp}$ \footnotemark[2]
and Vijay Shenoy $^{\S}$}

\footnotetext[1]{Current address: Division of Engineering and
Applied Sciences, Harvard University, USA} \footnotetext[2]{To
whom correspondence should be addressed; ({\em
lm@deas.harvard.edu, mkc4@lehigh.edu})}

\affiliation{$^{\sharp}$ Department of Chemical Engineering,
Lehigh University,
Bethlehem, PA 18015, USA. \\
$^{\star}$Department of Applied Mathematics and Theoretical
Physics, University of Cambridge, Wilberforce Road, CB3 0WA, UK. \\
$^{\S}$ Materials Research Centre, Indian Institute of Science,
Bangalore 560 012 India}

\begin{abstract}

Inspired by the observation that many naturally occurring
adhesives arise as textured thin films, we consider the
displacement controlled peeling of a flexible plate from an
incision-patterned thin adhesive elastic layer.  We find that
crack initiation from an incision on the film occurs at a load
much higher than that required to propagate it on a smooth
adhesive surface; multiple incisions thus cause the crack to
propagate intermittently. Microscopically, this  mode of crack
initiation and propagation in geometrically confined thin adhesive
films is related to the nucleation of cavitation bubbles behind
the incision  which must grow and coalesce before a viable crack
propagates. Our theoretical analysis allows us to rationalize
these experimental observations qualitatively and quantitatively
and suggests a simple design criterion for increasing the
interfacial fracture toughness of adhesive films.

\end{abstract}




\maketitle

\section{Introduction}
\begin{figure}[!htbp]
\centering
\includegraphics[height=8.5 cm]{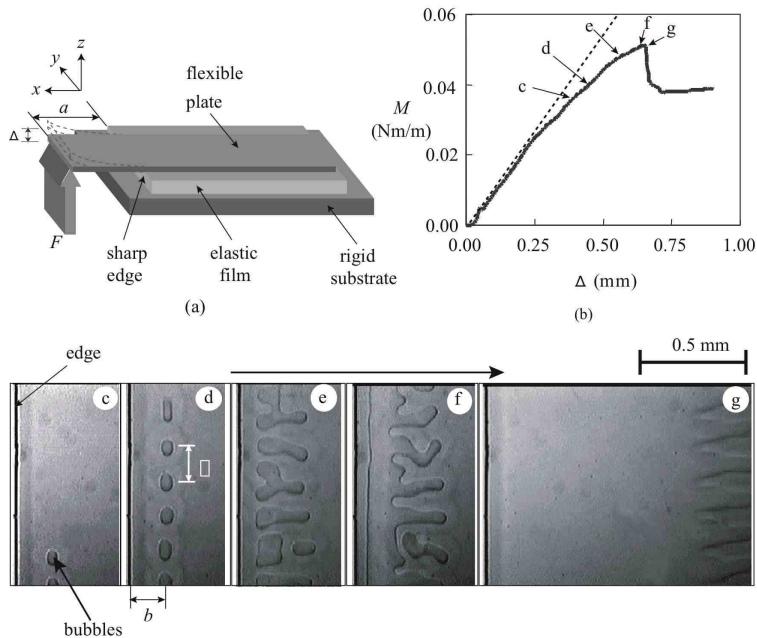}
\caption{(a) A flexible plate  in contact with an elastic
adhesive film is peeled off in a displacement-controlled
experiment. The film which remains bonded to a rigid substrate has
a sharp edge that serves to initiate a crack.  (b) The peeling or
lift-off moment, $M = Fa$, where $F$ is the load per unit width of
the plate, and $a$ is the  moment arm, as a function of the plate displacement
$\Delta$  shows a linear increase in the load (dashed line),
followed by a softening associated with the formation and growth
of cavitation bubbles before the sudden catastrophic drop once the
bubbles coalesce and the crack propagates on the smooth adhesive
film. (c)-(g) Video micrographs depict the sequence of
morphology changes in the contact zone as the peeling force is
increased. The images correspond to a film of thickness $h = 40
\mu$m and shear modulus, $\mu = 1.0$ MPa and a plate of rigidity
$D = 0.02$ Nm. We see that the film separates from the plate along
a series of cavitation bubbles at a distance $b$ behind the edge,
which  coalesce before the entire crack moves.}
\label{fig:fig1}
\end{figure}

The propagation of a crack on a smooth thin layer of adhesive
(Kaelble 1965, Gent \textit{et al.} 1975, Kendall 1975 \& Kinloch
\textit{et al.} 1994) has been extensively studied using peel
experiments. Although this model problem is of much relevance in
understanding the mechanical behavior of artificial adhesives,
naturally occurring adhesive surfaces in animals and insects in
particular present complex structural morphologies to alter the
physics of adhesion (Scherge and Gorb 2001). To explore the
mechanisms of crack initiation and propagation on these textured
interfaces,  we study the initiation propagation of cracks on
model adhesive layers which are patterned by sharp cuts and
discontinuities using a simple cantilever plate peeling experiment
(see Fig. \ref{fig:fig1}a). A flexible plate, in contact with a
thin film of adhesive which remains strongly bonded to a rigid
substrate, is lifted to initiate a crack from the edge of the
adhesive and the load-displacement curve is monitored to quantify
the force and energy required for crack initiation.

\section{Experiment}

We use thin films of polydimethylsiloxane (PDMS) as a model
adhesive film in our experiments. These crosslinked
elastomeric films of shear modulii between 0.2 - 3.1 MPa are
prepared by following the procedure described in Ghatak  and Chaudhury 2003. JKR (Johnson, Kendall and Roberts, 1971) contact
mechanics experiments of the networks indicate that they are
purely elastic and exhibit no hysteresis. The film is strongly
adhered to a rigid substrate which is attached to the stage of a
microbalance. Microscope cover slips coated with self assembled
monolayer (SAM) of Hexadecyltrichloro silane (HC),
hexafluorodecyltrichlorosilane (FC) molecules and with long chain
(MW $\sim 20000$) PDMS molecules  which can be brought into and out of contact
with the adhesive layer are used to explore the mechanics of
peeling.  The glass plate and the adhesive elastic film are first
rinsed thoroughly in de-ionized water to remove any static charge
and blow dried in nitrogen gas. We use a sharp razor blade to
make periodic incisions in the adhesive film   (Fig. \ref{fig:fig1}a) that are used to
arrest cracks and provide a barrier for their nucleation. We then
bring the flexible plate in complete contact with the elastic film
and keep it so for thirty minutes before quasi-statically lifting
the plate  at a distance away from the edge (Fig. 1 a) using a
micromanipulator that allows us to measure the vertical
displacement of the plate end $\Delta$, while the microbalance
allows us to simultaneously monitor the load $F$. During this
entire process, we view the contact zone near the edge of the film
using an optical microscope equipped with a CCD video camera to
study the morphology of adhesion during the process of crack
initiation.

{\bf Crack initiation at a single discontinuity:} In Fig.
\ref{fig:fig1}c-g we show a typical sequence of events leading to
crack initiation when a glass plate is lifted at a slow but
constant rate from the adhesive film. Initially the load
increases linearly with the displacement ($\Delta$) of the plate.
As the displacement is increased beyond a threshold, the moment
increases sub-linearly; this process is accompanied by the
nucleation of a series of cavitation bubbles (Fig.
\ref{fig:fig1} c) behind the edge. Although all the bubbles do not
all nucleate at the same time (Fig. \ref{fig:fig1}c - d) as
$\Delta$ is increased, they are periodically spaced with a
well-defined separation wavelength $\lambda$. Once this array of
bubbles is established behind the edge, a further increase in
$\Delta$ causes the existing bubbles to grow and coalesce (Fig.
\ref{fig:fig1}e - f) until they eventually reach the edge of the
adhesive film (Fig. \ref{fig:fig1}g). The almost straight crack
that results moves and reaches an equilibrium position determined
by the competition between plate bending and film shearing in a
displacement controlled experiment. If the lift-off displacement
is now reduced gradually this sequence is played out in reverse
and the crack closes as the plate reattaches to the film
everywhere except along a few uniformly spaced bubbles close to
the edge. Over relatively long times, the bubbles disappear and
homogeneous contact between the cover plate and the adhesive
film results. In this confined system,
cavitation does not occur randomly as thought previously (Gent
\textit{et al.} 1958, Kaelble 1971 \& Lakrout \textit{et al.}
1999); but instead is a manifestation of an adhesion induced
instabilities (Ghatak \textit{et al.} 2003) that leads to the
formation of the periodically spaced bubbles with a wavelength
$4h$  (Ghatak \textit{et al.} 2000).

\begin{figure}
\centering
\includegraphics[height=4.5cm]{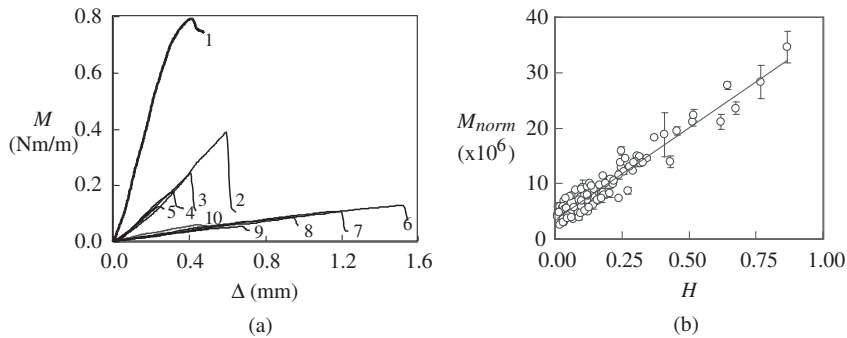}
\caption{(a) The peeling moment $M$ as a function of the plate displacement $\Delta$ for films made of a material with shear modulus $\mu = 0.9$ MPa but different thickness ($h = 40-800 \mu$m) and using cover plates of varying rigidities ($D = 0.02-1.35$ Nm). Curve 1:  $D = 1.35$ Nm, $h = 203 \mu$m. Curves 2 to 5: $D = 0.2$ Nm, $h = 560, 380, 110, 40 \mu$m. Curves 6 to 10: $D = 0.02$ Nm, $h = 760, 500, 300, 110, 40 \mu$m. (b) The maximum moment corresponding to crack initiation $M_{max}$ in (a) is rescaled to $M_{norm} = M_{max} \left(12 \mu/D \right)^{2/3}/Pi$ and plotted as a function of dimensionless film thickness $H = h/\left(D/12 \mu\right)^{1/3}$. The data collapses
on a single line given by $M_{norm} = 0.2 \times 10^{6} H + 0.02$. The non-zero vertical intercept when $H \rightarrow 0$ corresponds to the moment required for crack initiation and propagation on a
smooth rigid surface.} \label{fig:fig2}
\end{figure}

\begin{figure}
\centering
\includegraphics[height=4.5cm]{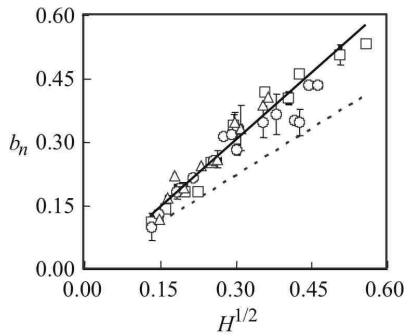}
\caption{The normalized bubble nucleation distance $b_{n} =
b/\left(D/12 \mu\right)^{1/3}$ as a function of $H^{1/2}$. Symbols
$\circ$, $\Box$ and $\triangle$ correspond to cover plates coated
with monomolecular-layer of HC, FC and PDMS respectively. The
solid line is the best fit for HC coated surfaces with $b_{n} =
1.1 H^{1/2} - 0.014$ (for FC and PDMS the fits are
indistinguishable from that for HC), while the dashed line is the
theoretical prediction (\ref{eq:eq20}): $b_{n} = 0.74 H^{1/2}$.}
\label{fig:fig3}
\end{figure}

Fig. \ref{fig:fig2} shows the peeling moment, $M = Fa$ as a
function of the displacement of the plate $\Delta$ for a variety
of film thicknesses ($h = 40 - 1000 \mu$m) and cover plate
rigidities ($D = 0.02, 0.2, 1.35$ Nm). For each cover plate, the
peeling moment $M$ increases with displacement $\Delta$ until $M =
M_{max}$, the maximum peeling moment corresponding to crack
initiation. A further increase in $\Delta$ causes $M$ to drop
abruptly to a much lower value as the crack then proceeds to
propagate on the smooth adhesive film. In Fig.
\ref{fig:fig2}b), we plot the scaled maximum moment  $ M_{norm} =
M_{max} \left(12 \mu/D \right)^{2/3}/\Pi$ ($\Pi= A/\left (6\pi
{d_c}^3 \right )$ is the van der Waals stress associated with a
separation distance $d_c$, $A$ being the Hamaker constant between
the two surfaces) as a function of the  scaled thickness $H =
h/\left(D/12 \mu\right)^{1/3}$  and see that $M_{norm} $ varies
linearly with $H$.  Similar linear relationships persist when the
cover plates are coated with other self-assembled monolayers such
as those made of FC   and PDMS.  In Fig. \ref{fig:fig3}, we show
that the distance of the cavitation bubbles from the edge of the
film follows the scaling law $b \sim h^{1/2}$.

\begin{figure}
\centering
\includegraphics[height=8.5cm]{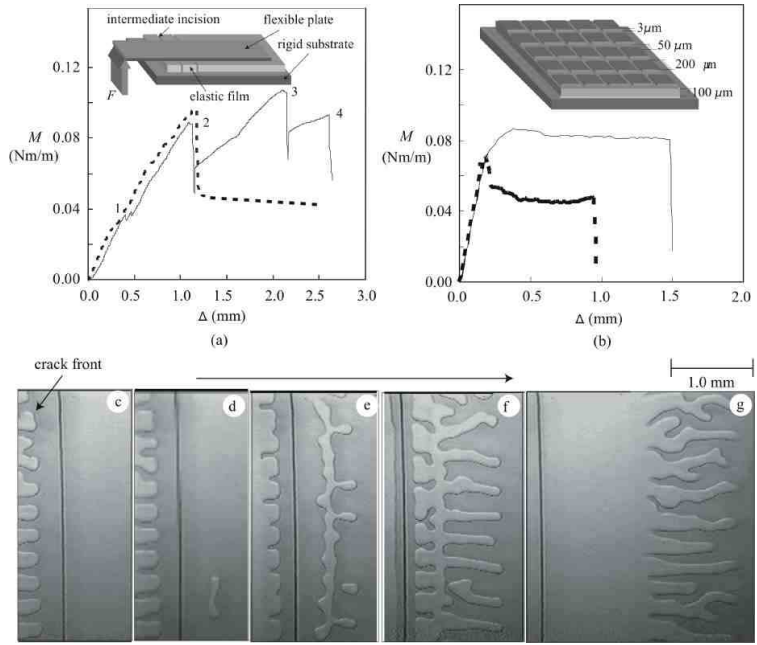}
\caption{(a) When the elastic film  has multiple incisions, the crack moves intermittently as the edges create crack initiation barriers. A typical plot of the peeling moment $M$ vs. the plate displacement $\Delta$ for such an experiment shows the presence of multiple peaks. The film thickness $h = 360 \mu$m. The peaks correspond to incisions 1,2,3 4 which are located at  20, 21, 27 and 33.5 mm  from the loaded end of the plate. The low crack initiation moment for incision 1 is due to defects. The dashed line depicts the moment-displacement characteristic in the absence of incisions 3 and 4. The upward
jump in $M$ following crack initiation is due to the rapidity with which the crack propagates; this makes it difficult to measure the intermediate values of the torque arm $a$ experimentally. (b) When the simple
incision-textured film is replaced by a chocolate-bar-textured elastic film which is prepared using a mould, the peeling moment does not indicate any intermittency. For comparison,  crack
propagation on a featureless smooth film leads to a moment displacement curve indicated with a dashed line. We see that texturing leads to a large enhancement of the interfacial fracture
toughness. Here, $D = 0.02$ Nm and $\mu = 0.9$ MPa, and the dimensions of the texture are as indicated. (c)-(g) Video-micrographs depict the sequence of crack initiation close to
an incision on a film of thickness $80 \mu$m. In all cases the shear modulus of the film $\mu =0.9$ MPa and the plate flexural rigidity $D = 0.02$ Nm.}
\label{fig:fig4}
\end{figure}
\begin{figure}
\centering
\includegraphics[height=5cm]{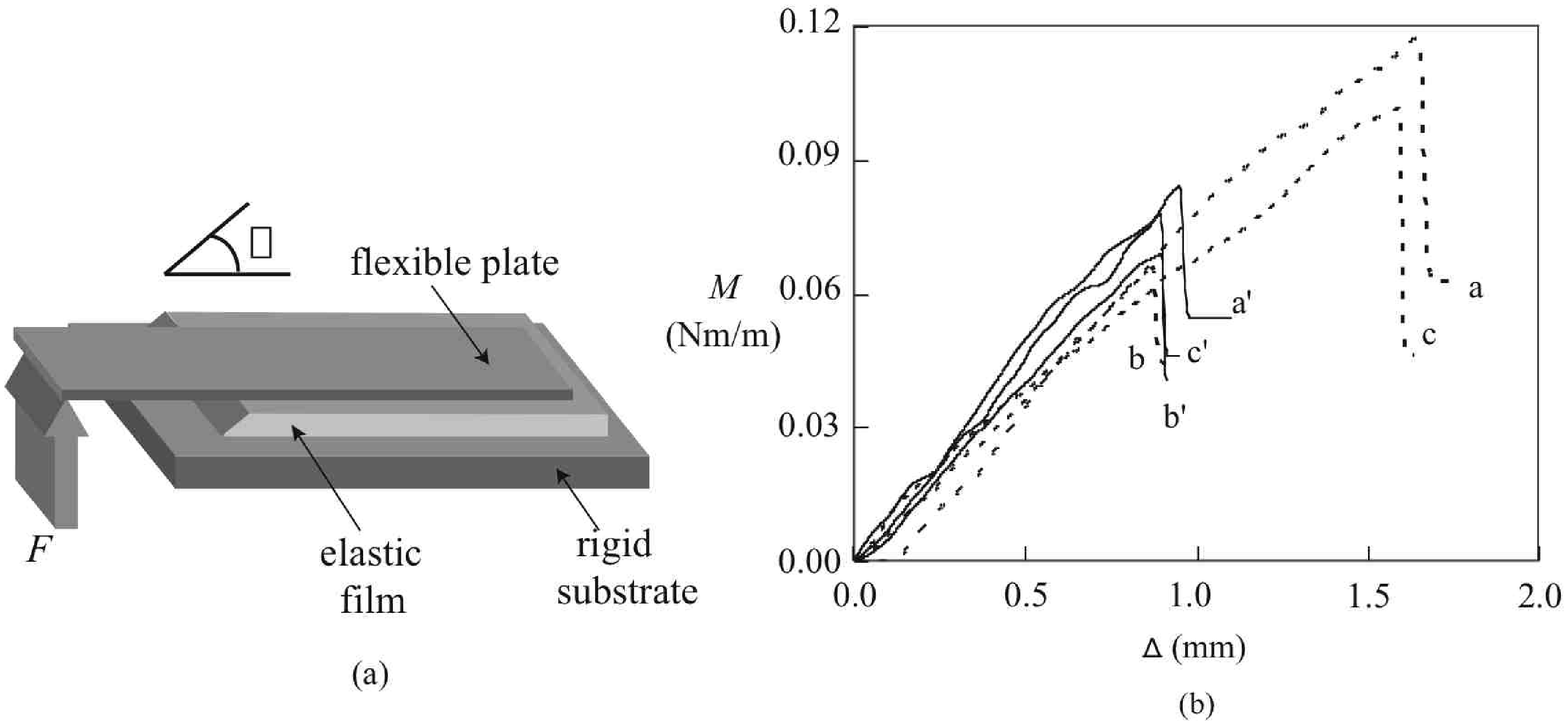}
\caption{(a) The effect of obliquity of the edge on the ease of
crack initiation is studied by cutting the {elastic} film
at different $\theta$. (b) The moment-displacement curves (a, a'),
(b, b') and (c, c') correspond to $\theta = 90^{o}, 48^{o}$ and
$132^{o}$ . The solid and dashed lines correspond to films of
thicknesses $h = 356$ (curves a', b' and c') and $762 \mu$m
(curves a, b and c) respectively. In all cases the shear modulus
of the film $\mu =0.9$ MPa and the plate flexural rigidity $D =
0.02$ Nm.} \label{fig:fig5}
\end{figure}

{\bf Crack Initiation at multiple discontinuities:} Similar
experiments were also carried out on elastic films with multiple
parallel incisions ($5-10$ mm apart) which span the width of the
film. In Fig. 4a, we show  a typical load-displacement curve for
this situation and see the signature of stick slip behavior. Video
micrographs in Fig. \ref{fig:fig4}c-g show the sequential crack
arrest and initiation on such an incision patterned film. Here,
the crack initiates from one such incision at a sufficiently high
load but gets arrested at the next one. Surprisingly, the crack
front stops before it reaches the next incision (Fig.
\ref{fig:fig4}c) and remains there while bubbles nucleates on the
opposite side (Fig. \ref{fig:fig4}d-f) of the incision. Finally,
the two fronts meet at the incision (Fig. \ref{fig:fig4}g)
forming a crack that then propagates rapidly.  When the experiment
is repeated on a two-dimensional textured and patterned surface
prepared by using a mould (inset of Fig. \ref{fig:fig4}b)
multiple crack arrest and initiation events lead to a high peeling
moment (solid line) which remains constant and does not show the
stick-slip like characteristics for incisions that are widely
separated. As we will see later, the disappearance of the
stick-slip behaviour is associated with the fact that the
characteristic length scale of the texture is smaller than a
critical threshold. In such a situation, since the peeling moment
is much higher than that on a smooth adhesive film, we
see that the fracture toughness of the interface is also
significantly enhanced by  texturing, quite contrary to normal
intuition.

{\bf Obliquity of the edge:} To understand the role of the edge in
crack initiation, we also carried out experiments on films in
which the  incision makes an angle $\theta$ different from
$90^{o}$ (Fig. \ref{fig:fig5}a). In Fig. \ref{fig:fig5}b we
plot the peeling moment as a function of displacement $\Delta$ for
various $\theta$ and see  that changes in $\theta$ affect the
crack initiation moment more significantly for  thick films  than
for thin ones. As $\theta \rightarrow 0$ the critical peeling
moment becomes smaller and smaller, consistent with the fact that
when $\theta = 0$ there is no edge at all so that there is no
barrier for crack initiation, and suggests that $\theta$ is a
measure of crack blunting. Further experiments are needed to
quantify this effect and will be subject of future study.

\section{Theory}

We now turn to an approximate theory to rationalize our
experimental findings. Assuming that the adhesive film is
incompressible, linearly elastic and loaded in plane strain, the
equations of equilibrium are
\begin{eqnarray}\label{eq:eq4}
 P_{x} &=& \mu \left ( u_{xx} + u_{zz} \right ),\nonumber \\
 P_{z} &=& \mu \left ( w_{xx} + w_{zz} \right )
\end{eqnarray}
Here and elsewhere $a_{b} = \frac{\partial a}{\partial b}$,
$P(x,z)$ is the pressure in the elastic film,
$u\left(x,z\right)$ and $w\left(x,z\right)$ are the components of
displacement field in the $x$ and $z$ directions (Fig. 1 a), and
$\mu$ is the shear modulus of the adhesive. The pressure itself is
determined by the constraint of incompressibility, which in its
linearized form can be written as
\begin{eqnarray}\label{eq:eq5}
 u_{x} + w_{z} = 0
\end{eqnarray}
With the origin of coordinate system at the corner of the incision in the film ,  as shown in Fig. \ref{fig:fig1}a, the corresponding boundary conditions (b.c.) are
\begin{eqnarray}\label{eq:eq6}
& u \left(x, 0 \right) = 0, & w \left (x, 0 \right ) = 0  \nonumber \\
& u \left(x, h\right) = 0, & P = D \xi_{xxxx} \hspace
{5 mm}
\text{for} \hspace {5 mm} x<0 \nonumber \\
& ~~ & P = 0 \hspace {5 mm} \text{for}
\hspace {5 mm} 0<x<a
\end{eqnarray}
Here $\xi(x) = w(x,h)$ is the vertical displacement of film at $z = h$, and $a$ is the distance of the line of application of the peeling force from the contact line $x=0$. For a thin film with a
large lateral length scale $L$, vertical gradients in the displacement fields are much larger than
horizontal gradients , so that we may use the lubrication approximation (Batchelor, 1967). Then $u_{xx}
\sim U/L^2 \ll U/h^2$, $w_{xx} \sim hU/L^3 \ll U/h^2$ and $w_{zz} \sim U/\left(Lh\right) \ll U/h^2$ and the equations of equilibrium (\ref{eq:eq4}) simplify to
\begin{eqnarray}\label{eq:eq7}
 P_{x} &=& \mu u_{zz},\nonumber \\
  P_{z} &=& 0.
 \end{eqnarray}
Integrating   (\ref{eq:eq7}-\ref{eq:eq6}) yields
\begin{eqnarray}\label{eq:eq8}
 u \left (x, z \right ) & = & \frac{D}{2 \mu} \xi_{xxxxx} \left ( z^{2} - h z \right ) \label{disp}
\end{eqnarray}
Substituting (\ref{disp}) into the depth-integrated continuity equation (\ref{eq:eq5})  and linearizing the result leads to an equation for the vertical displacement of the cover plate in the region $x<0$ where it is attached to the adhesive film
\begin{eqnarray}\label{eq:eq11}
 \xi_{xxxxxx} - \frac{12 \mu}{D h^{3}} \xi = 0 \hspace {5 mm} \text{for} \hspace {5 mm} x<0
 \end{eqnarray}
In the region $0<x<a$ where the film is not in contact with the plate, the displacement of the cover plate satisfies
\begin{eqnarray}
\label{eq:eq11b}
 \xi_{xxxx} = 0 \hspace {5 mm} \text{for} \hspace {5 mm} 0<x<a
 \end{eqnarray}
Since the film must be flat far away  from  the contact line
\begin{eqnarray}\label{eq:eq12}
\left. \xi \right |_{x \rightarrow -\infty} = 0, \hspace{5 mm}
\left. \xi_{x} \right |_{x \rightarrow
-\infty} = 0, & \left. \xi_{xx} \right |_{x \rightarrow -\infty} = 0.
\end{eqnarray}
Continuity of the displacement, slope, bending moment, vertical
shear force and the pressure at the contact line imply that
\begin{equation}\label{eq:eq13}
\begin{array}{lll}
\left. \xi |_{0-} = \xi \right|_{0+}, &
\left. \xi_x \right |_{0-} = \left. \xi_x \right  |_{0+} & \\
\left.  \xi_{xx} \right |_{0-} =
\left.  \xi_{xx} \right |_{0+}, &
\left. \xi_{xxx} \right |_{0-} =
\left.  \xi_{xxx} \right |_{0+},
\left.  \xi_{xxxx} \right |_{0} =0.
\end{array}
\end{equation}
Finally, at $x=a$, where the flexible plate is freely pivoted
while being lifted vertically by an amount $\Delta$, the boundary
conditions are
\begin{eqnarray}\label{eq:eq14}
\left. \xi \right|_{x=a} = \Delta, & \left.  \xi_{xx} \right |_{x = a} = 0.
\end{eqnarray}
We pause briefly to consider the assumptions inherent in our
approach. Lubrication theory clearly must break down over a length
scale of order $O(h)$ from the edge. However, if we consider
variations over scales much larger than $h$, these edge effects
can be safely ignored. As we shall see later, this is indeed the
case. The edge of the incision  where the plate first loses
contact with the elastic film acts to pin the contact line. This
allows us to specify the deflection of the plate $\Delta$ and the
distance of the contact line from the point of application of the
force $a$ independently. This is in contrast to the case of a
flexible plate in contact with a smooth elastic film, where it is
not possible to specify both $a$ and $\Delta$ since there is a
relation between them in terms of the work of adhesion.

Solving (\ref{eq:eq11}) subject to (\ref{eq:eq12}-\ref{eq:eq14})
for the interfacial displacement $\xi\left(x\right)$ leads to
\begin{eqnarray}\label{eq:eq16}
\xi \left(x \right ) = \left \{%
\begin{array}{l}  F' e^{kx/2}\left ( e^{kx/2} + \frac{2 + ak}{1 + ak} \cos \left (\frac{\sqrt{3}kx}{2} \right ) +
 \frac{1}{\sqrt{3}} \frac{ak}{1+ak} \sin \left(\frac{\sqrt{3}kx}{2} \right ) \right ), \hspace{2 mm}
\textrm{$x < 0$} \\
F' \left ( \frac{3 + 2ak}{1 + ak} + 2 kx + \right.
\left. \frac{ak}{1 + ak} \frac{\left (kx \right)^2}{2} - \frac{1}{1 + ak} \frac{\left(kx \right )^3}{6} \right ), \hspace{2
mm} \textrm{$ 0 < x < a$}
\end{array}%
\right.
\end{eqnarray}
where $F' = 3\Delta\left (1 + ak \right)/ \left (
(ak)^3+6(ak)^2+12(ak)+9 \right)$ and $k^{-1}= \left( D h^{3}/12
\mu \right )^{1/6}$ are two characteristic lengths that arise
naturally in the problem. The scale $k^{-1}$ determines the lateral extent of the film over which the peeling deformation is felt;  on scales larger than $k^{-1}$ (\ref{eq:eq16}) shows that the interface displacement decays exponentially.  For  typical
experimental parameter values,  $kh = (12 \mu h^3/D)^{1/3} \sim
(10^6 . 10^{-14}/10^{-2})^{1/3} \sim 10^{-2} <<1$, so that crack tip and edge effects are relatively small.  Using (\ref{eq:eq11}) we can now determine the normal traction on the surface of the film
which, in the lubrication approximation, is given by $\left.
\sigma_{33}\right |_{h} = -P + 2\mu w_{z} \sim -P=D \xi_{xxxx}$.
For long wavelength deformations $ak >> 1$ so that
\begin{eqnarray}\label{eq:eq19}
  \sigma_{33}|_{z=h} =-P \approx  -Fak^{2} e^{kx/2} \left( e^{kx/2}-
 \cos \left(\frac{\sqrt{3}kx}{2}\right) +  \frac{1}{\sqrt{3}} \sin \left(\frac{\sqrt{3}kx}{2}\right ) \right )
\end{eqnarray}
where $\left. F= -D \xi_{xxx} \right |_{x=a}$. In Fig.
\ref{fig:fig5} we show the variation of the vertical displacement
$ w\left(x, h \right ) = \xi$ and the normal traction
$\sigma_{33}(x,h)$.  We note that both the displacement and the
traction are oscillatory with exponentially decaying amplitudes;
in particular the normal traction has a negative maximum at a
distance $b$ from the edge, given by
\begin{eqnarray}\label{eq:eq20}
 b = 0.74 \left(\frac{Dh^3}{12\mu}\right)^{1/6},
\end{eqnarray}
where the pressure
\begin{eqnarray}\label{eq:eq21}
 P = -  \left . \sigma_{33}\right |_{max} = - 0.32\left(\frac{12 \mu}{D} \right)^{1/3} \left(\frac{Fa}{h}\right).
\end{eqnarray}
This large tensile traction at some distance behind the contact line rationalizes our observations of cavitation bubbles. However  the characteristic periodicity of the bubbles in the transverse direction (parallel to the contact line) with a wavelength $\lambda \approx 4h$ (Ghatak and Chaudhury, 2003) requires a three-dimensional stability analysis of  the planar solution and is beyond the scope of the current study. Here the role of the incisions is to pin the contact line and prevent the crack from being initiated until a threshold stress is reached, thus raising the effective toughness of the interface. Our experimental observations of the dimensionless cavitation bubble nucleation distance $b_n$ shown in Fig. 3 are qualitatively consistent with (\ref{eq:eq20})  although quantitatively there is a discrepancy of about $25\%$. This  could be due to the use of the particular boundary condition that the pressure is continuous at the contact line $x=0$. If we replace this by a different  boundary condition   $\left. \xi \right|_{x=0-}=\xi_{x=0+} = \delta$, and calculate $\delta$ by minimizing the total energy of the system, we get $b_n \approx 1.0$, in accordance with our experimental measurements. In fact,
the actual condition at the contact line is determined by the details of the microscopic interaction between the two surfaces and is probably somewhat intermediate between these two cases.

\begin{figure}
\centering
\includegraphics[height=4.5cm]{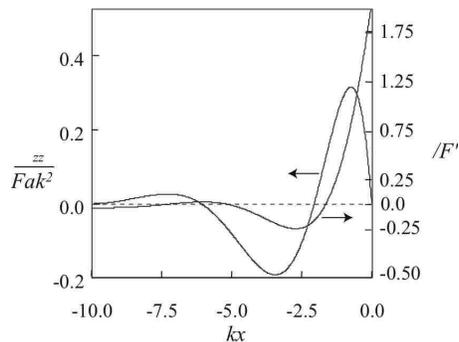}
\caption{The dimensionless normal traction
$\sigma_{zz}/Fak^{2}=-P(x,h)/Fak^2$ and the dimensionless
displacement $w(x,h) = \xi/F'$ at the film interface as a
function of dimensionless distance $kx$ from the edge are
oscillatory exponentials. The dimensionless maximum tensile stress $\Sigma$ occurs at a
dimensionless distance $b_n$ behind the incision.}
\label{fig:fig6}
\end{figure}

The exponentially decaying stress profile (Fig. \ref{fig:fig6})
also explains why the crack gets arrested before it reaches the an
incision. Since the normal traction vanishes at the incision, the
next crack is  initiated via bubble nucleation and coalescence on
the other side of the incision and the whole scenario repeats
itself. When the distance between incisions becomes of the order
of or less than the characteristic stress decay length $k^{-1} =
\left(Dh^3/\mu\right)^{1/6}$ the crack feels the effect of
incisions continuously, and the intermittent behavior of the
peeling moment is replaced by a much higher constant value for a
finely textured surface (Fig. \ref{fig:fig4}(b)).

To obtain the stress associated with bubble nucleation, we rearrange equation (\ref{eq:eq21})  so that
\begin{eqnarray}\label{eq:eq22}
 M_{max} = F_{max}a = 3\sigma_{c} h \left(\frac{D}{12\mu} \right)^{1/3}
\end{eqnarray}
Here $\sigma_{c}$ is the critical stress associated with bubble
nucleation, which can be determined by comparing (\ref{eq:eq22})
with our experimental data and yields $\sigma_{c} \sim 6 \times
10^{4}$ N/m$^{2}$. For comparison, with a Hamaker constant $A \sim
4 \times 10^{-19}$ J, a separation distance $d_{c} \sim 1.5$ \AA
\hspace{1mm} the van der Waals pressure, $\Pi = A/\left(6\pi
{d_{c}}^{3}\right)  \sim 6 \times 10^{9} $ N/m$^{2}$. The
experimentally obtained low value of $\sigma_{c}$ suggests that
the two surfaces do not remain in perfect contact and are
separated by an average distance of  $ \sim 20$ \AA
possibly due to the intrinsic roughness of the adhesive films.
However, our experiments even with films of very low
root-mean-square roughness ($\sim 3$ \AA) results in a low critical
stress ($\sigma_c \sim 6 \times 10^{4}$ N/m$^{2}$), signifying
that other factors may be responsible as well.

\section{Discussion}

In this note, we have demonstrated the qualitative difference between crack initiation and crack propagation in the context of peeling a flexible plate from a thin textured adhesive film. The origin of this difference may be ascribed to the formation of  cavitation bubbles behind the contact line which increases the load required for peeling by creating a convoluted crack front.  A different way of enhancing the load for crack initiation is to use
films with oblique incisions, and leads to effective way of blunting or sharpening the crack tip.
A simple theory allows us to explain these observations qualitatively and quantitatively, and leads to a design criterion for enhancing the interfacial fracture toughness of a flexible plate in contact with an adhesive film: the pattern has to be microstructured on a length scale smaller than or equal to
the stress decay length $(Dh^3/\mu)^{1/6}$.

We finally return to our motivation of biological attachment devices (Scherge \textit{et al.} 2001) which show a variety of textured contact surfaces. Our experiments on model patterned systems suggest that the enhanced fracture toughness in these biological settings is a rather subtle effect owing to the
difference between crack initiation and propagation on a patterned surface. Multiple crack arrest and initiation on these surfaces results in the dissipation of the elastic energy in much the same
way as  for fracture of soft elastomers (Lake and Thomas 1967): even if all the polymers in the film are strained, when a bond ruptures the broken parts relax under zero load leading to
dissipation of energy. Nature seems to have taken advantage of these principles in designing the attachment pads of insects and other sticky surfaces for millenia, and so all that remains is for
us to understand and  mimic her infinite variety.

\begin{acknowledgements}
We gratefully acknowledge discussions with M. Argentina. LM and MKC acknowledge the support of the US Office of Naval Research. \\

\end{acknowledgements}

\end{document}